\documentstyle[twocolumn,prl,aps,amsfonts]{revtex}
\newcommand{\rmD}{\mbox{\scriptsize D}}
\newcommand{\rmSI}{\mbox{\scriptsize SI}}
\newcommand{\RR}{{\Bbb R}}
\begin{document}
\title{
Comment on ``Superinstantons and the Reliability of Perturbation Theory
in Non-Abelian Models"
}
\maketitle
{
In a recent letter \cite{PS1}
Patrascioiu and Seiler argued that the results of standard
perturbation theory (PT) are not valid for two dimensional models with a
non-abelian global symmetry.
They considered such a theory in a finite box of size $L$,
with special boundary conditions (BC) called ``superinstantons" (SI)
and showed that in
the thermodynamic limit $L\to\infty$ the 2-loop corrections are finite, but
differ from those obtained from PT with standard BC.
They concluded that, when SI configurations are taken into
account, renormalization group (RG) $\beta$-functions are modified and
that the limit $L\to\infty$ and the weak-coupling expansion do not commute.
\par
In fact the results of \cite{PS1} do not contradict standard PT.
Indeed one can show from general principles that:
(1) PT with a SI BC is infra-red (IR) divergent;
(2) the IR divergent terms are associated, via the short distance operator
product expansion (OPE), to singular local operators, not present for classical
backgrounds;
(3) taking into account these operators, the perturbative RG functions are not
modified.
\par
Let us show this for the non-linear O($N$) $\sigma$-model considered in
\cite{PS1},
with $N$-component unit vector field
$\vec S=(\vec\pi,\sigma=\sqrt{1-\vec\pi^2})$,
defined in a
square box $\Lambda_L=[-L/2,L/2]\times[-L/2,L/2]$, with Dirichlet (D) BC
$\vec\pi=\vec 0$ on $\partial\Lambda_L$, and with the additional SI constraint
$\vec\pi(0)=\vec 0$ at the origin.
The 2-point function with the SI~BC is related to that with the D~BC by
\begin{equation}
\langle\vec S(x)\cdot\vec S(y)\rangle_{\rmSI}=
{\langle\vec S(x)\cdot\vec S(y)\delta(\vec\pi(0))\rangle_{\rmD}
\over
\langle\delta(\vec\pi(0))\rangle_{\rmD}}
\end{equation}
with $\delta(\vec\pi)$ the Dirac distribution in $\RR^{N-1}$.
Its perturbative expansion is
$\langle\vec S(x)\cdot\vec S(y)\rangle_{\rmSI}=1+c_1 g +c_2 g^2 +\cdots$
For simplicity we first regularize the short-distance divergences by
using dimensional regularization, with space dimension
$D=2-\epsilon$ ($\epsilon>0$), and use the continuum action
$S={1\over 2g}\int d^Dx(\partial\vec S(x))^2$.
Adapting the results of \cite{David},
and the techniques of \cite{DDG,LasLip} to deal with the singular operator
$\delta(\vec\pi)$,
it is easy to show that
the $L\to\infty$ expansion of (1) is given by a sum over local
operators $A(0)$ located at $x=0$, and with support in field space at
$\{\vec\pi(0)=0\}$
\begin{equation}
\langle\vec S(x)\cdot\vec S(y)\rangle_{\rmSI}=
\sum_{A} C^A(x,y)
{\langle A(0)\rangle_{\rm D}\over\langle\delta(\vec\pi(0))\rangle_{\rmD}}
\end{equation}
where the OPE coefficients $C^A$ are {\it independent} from the specific BC
used
and from $L$ and are {\it finite} in PT.
The BC and $L$ dependence is entirely contained in the expectation values
ratios (evaluated in the box $\Lambda_L$ with D BC)
$\langle A(0)\rangle_{\rmD}\big/\langle\delta(\vec\pi(0))\rangle_{\rmD}$,
which scale as
$L^{-{\mbox{\scriptsize dim}}[A]} f_A(gL^\epsilon)$,
with $f_A$ calculable in PT, and where $\mbox{dim}[A]$ is the canonical
dimension of $A$ ($\mbox{dim}[\vec\pi]=0$, $\mbox{dim}[x]=-1$).
Only operators with $\mbox{dim}[A]=0$ can give finite or divergent
contributions when $L\to\infty$,
operators with $\mbox{dim}[A]>0$ give subdominant corrections.
The first dimensionless operator is $A_0=\delta(\vec\pi)$:
$f_{A_0}=1$ and the coefficient $C^{A_0}=1+c^{\scriptscriptstyle (0)}_1 g+
c^{\scriptscriptstyle (0)}_2 g^2 +\cdots$ gives the standard IR finite PT.
However additional operators appear, of the form
$A_n=(\Delta_{\vec\pi})^n\delta(\vec\pi)$, with $\Delta_{\vec\pi}$ the
Laplacian in $\RR^{N-1}$.
The first one, $A_1$, is such that $C^{A_1}=c^{\scriptscriptstyle (1)}_2 g^2+
c^{\scriptscriptstyle (1)}_3 g^3+\cdots$, and that $f_{A_1}(g)=
f^{\scriptscriptstyle (1)}_{-1}g^{-1}+f^{\scriptscriptstyle (1)}_{0}+
f^{\scriptscriptstyle (1)}_1 g+\cdots$
Simple diagrammatics shows that $C^{A_n}={\cal O}(g^{2n})$ and
$f_{A_n}={\cal O}(g^{-n})$.
Therefore the coefficient of $g$ of (1) behaves as
$c_1=c^{\scriptscriptstyle (0)}_1+L^{-\epsilon}c^{\scriptscriptstyle (1)}_2
f^{\scriptscriptstyle (1)}_{-1}+ \cdots$ when $L\to\infty$.
Its IR limit coincides with the standard PT result,
$c^{\scriptscriptstyle (0)}_1$.
The coefficient of $g^2$ behaves as
$c_2=c^{\scriptscriptstyle (0)}_2+
c^{\scriptscriptstyle (1)}_2 f^{\scriptscriptstyle (1)}_0+
L^{-\epsilon}c^{\scriptscriptstyle (1)}_3 f^{\scriptscriptstyle (1)}_{-1}+
\cdots$
and has a finite IR limit, different from the PT result,
$c^{\scriptscriptstyle (0)}_2$.
However, the coefficient of $g^3$ is IR singular
$c_3=c^{\scriptscriptstyle (1)}_2 f^{\scriptscriptstyle (1)}_1
L^{\epsilon}+\cdots$,
as well as the higher order terms.
\par
These conclusions are independent of the regularization, and can be extended to
the $D=2$ lattice model of \cite{PS1}:
with SI BC, the 2-point function is IR finite at order $g^2$
but differs from standard PT; at order $g^n$, $n>2$, it is IR divergent, with
$\ln(L)^{n-2}$ terms.
These results are valid {\it order by order} in PT, and apply to the
renormalized theory as well: to construct the continuum limit for finite $L$,
besides the coupling constant and wave-function renormalization,
one must also renormalize the SI insertion operator $A_0(0)$.
Taking this effect into account, the PT $\beta$-functions are unchanged, but
the renormalized theory with SI BC is IR divergent at order $g^3$ and beyond.
Similar problems are expected to occur for the non-abelian gauge theories
considered in \cite{PS2}.
\par
\medskip\noindent
{\bf Fran\c cois DAVID}$^{(1)}$\par
CEA, Service de Physique Th\'eorique, CE-Saclay\par
F-91191 Gif-sur-Yvette Cedex, FRANCE\par
\medskip
\noindent $^{(1)}$Member of CNRS\par
}


\begin{references}
\bibitem{PS1} A. Patrascioiu and E. Seiler, Phys. Rev. Lett. {\bf 74} (1995)
1920-1923.
\bibitem{David} F. David, Commun. Math Phys. {\bf 81} (1981) 149.
\bibitem{DDG} F. David. B. Duplantier and E. Guitter, Nucl. Phys. {\bf B394}
(1993) 555.
\bibitem{LasLip} M. L\"assig and R. Lipowsky, Phys. Rev. Lett. {\bf 70} (1993)
1131.
\bibitem{PS2} A. Patrascioiu and E. Seiler, Phys. Rev. Lett. {\bf 74} (1995)
1924-1927.
\end{references}
\end{document}